\documentclass[prl,aps,amssymb,showpacs,twocolumn,amsmath,floatfix]{revtex4}
\usepackage{tabularx}
\usepackage{bm}
\usepackage{euscript}
\usepackage{epsfig,psfrag,subfigure}
\usepackage{graphicx}
\usepackage{color}
\usepackage{amsfonts}
\usepackage{exscale}
\usepackage{amsbsy}
\usepackage{hyperref}
\input{epsf}

\def\be{\begin{equation}}
\def\ee{\end{equation}}
\def\ba{\begin{eqnarray}}
\def\ea{\end{eqnarray}}

\def\SRO{\rm Sr$_2$RuO$_4$}

\begin{document}
\title{
Hidden quasi one-dimensional superconductivity in \SRO}
\author{S. Raghu, A. Kapitulnik, and S. A. Kivelson}
\affiliation{Department of Physics, Stanford University, Stanford, CA 94305}
\date{\today}

\begin{abstract}
Using an asymptotically exact weak coupling analysis of a multi-orbital Hubbard model of the electronic structure of \SRO, we show that the interplay between spin and charge fluctuations leads unequivocally to triplet pairing which originates in the quasi-one dimensional bands.  The resulting superconducting state spontaneously breaks time-reversal symmetry and is of the form $\Delta \sim \left(p_x + i p_y \right) \hat{z}$ with sharp gap minima and a d-vector that is only {\it weakly} pinned.  The superconductor is topologically {\it trivial} and hence lacks robust chiral Majorana fermion modes along the boundary.  The absence of topologically protected edge modes could explain the surprising absence of experimentally detectable edge currents in this system.  
\end{abstract}


\maketitle

{\it Introduction - }
{\rm Sr$_2$RuO$_4$} is a layered perovskite material, isostructural to the  hole-doped 214 family of  cuprate superconductors.
Below $T \sim 50 K$, it exhibits Fermi liquid behavior and  undergoes a superconducting transition at $T_c = 1.5 K$.  There is compelling experimental evidence which suggests that this superconducting state has odd parity\cite{Nelson2004, Kidwingira2006} and spontaneously breaks   time-reversal \cite{luke1998,Xia2006,Kidwingira2006} symmetry.  One of the simplest superconducting gap functions which meets both of  these requirements is the chiral p-wave state, $ \vec{\Delta}(\bm p)  \propto \left(p_x + i p_y \right) \hat{z}$, a quasi-two-dimensional version of superfluid $^3$He-A \cite{Anderson1975, Rice1995}. 

In its simplest form, this chiral pairing gives rise to a topological superconductor: all Bogoliubov quasiparticle excitations are gapped in the bulk  whereas  topologically protected chiral Majorana fermion modes exist at the edge of the system and
in vortex cores \cite{Read2000}.  These  modes are  
robust against all perturbations, 
including disorder,
so long as the BCS pairing gap in the bulk remains finite.  In addition, spontaneous supercurrents are expected at sample edges and domain walls 
\cite{matsumoto1999,stone2004}.

 However, scanning squid and Hall bar imaging studies \cite{bjornsson2005} have revealed that  edge currents of the expected magnitude are {\it not} found in \SRO.  Moreover, low temperature power laws in the electronic specific heat\cite{Nishizaki1999} and the nuclear spin relaxation $1/T_1$\cite{Ishida2000} suggest that 
 this material is not  a simple chiral superconductor, which would exhibit exponentially activated behavior in both of these quantities.  
 The  form of the superconducting order parameter which accounts for all of 
 the observed phenomena  remains unknown. Resolution of this puzzle could come from a careful consideration of the normal state properties, 
 which are known with unprecedented detail \cite{Bergemann2003, Mackenzie2003}.  

The Fermi surface of \SRO consists of 3 sheets, denoted $\alpha, \beta, \gamma$\cite{Bergemann2003, Mackenzie2003}.  The $\alpha$ and $ \beta$ sheets are hole and electron pockets respectively; they are comprised primarily of the Ru $d_{xz}, d_{yz}$ orbitals which form quasi one dimensional bands.  The $\gamma$ sheet is composed mainly of the Ru $d_{xy}$ orbital, which forms a quasi two dimensional band.   A variety of experiments have shown that the system behaves as a quasi two dimensional Fermi liquid with considerable effective mass enhancements\cite{Mackenzie2003}.  Therefore, it is likely that electron correlations play a significant role in influencing the pairing mechanism of this system.

In this paper, we present a microscopic theory of superconductivity in \SRO.  Using a simple extension of a recently developed asymptotically exact weak-coupling analysis of the Hubbard model\cite{Raghu2010}, we show that   the dominant superconducting instability is in the triplet channel and occurs on the quasi-1D Fermi surfaces of \SRO.   The resulting superconducting state spontaneously breaks time-reversal symmetry. It exhibits node-like behavior since it possesses points on the Fermi surface where the gap is  parametrically small. It supports Andreev bound states at domain walls and at the edges of the system.  However, it is a topologically trivial superconductor without  chiral Majorana  fermion modes, or detectable spontaneous supercurrents at the edges.

 {\it Microscopic model -} 
We consider a simple Hamiltonian 
with three bands derived from the Ru $t_{2g}$ orbitals  
\begin{figure}
\includegraphics[width=2.2in]{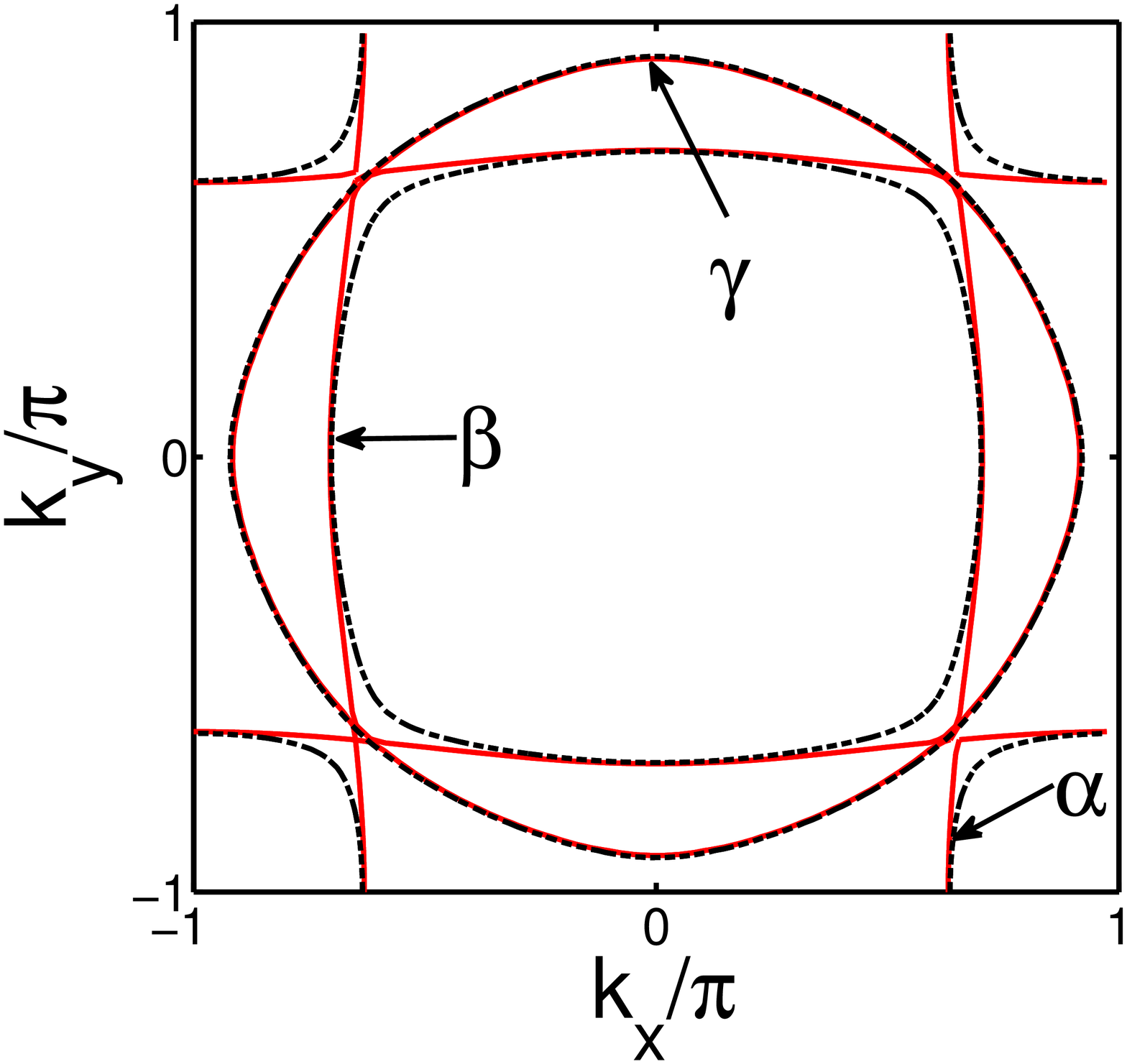}
\caption{(Color online) Tight-binding Fermi surface for the non-interacting Hamiltonian $H_0$.  Hybridizations among different orbitals are neglected in the sold curve and are included in the dashed curve.  }
\label{fs}
\end{figure} 
\begin{equation}
H = H_0 + U \sum_{i \alpha} n_{i \alpha \uparrow} n_{i \alpha \downarrow} + \frac{V}{2} \sum_{i, \alpha \ne \beta} n_{i \alpha} n_{i \beta} + \delta H
\end{equation}
Here, we introduce vector indices such that $\alpha =x$, $y$, and $z$, refer, respectively, to the Ru 
$d_{xz}$, $d_{yz}$ and $d_{xy} $ orbitals, $n_{i \alpha \sigma}$ is the density of electrons having spin $\sigma$ at position $i$ in orbital $\alpha$  and $n_{i \alpha} = \sum_{\sigma} n_{i \alpha \sigma}$.  The strength of the repulsive interaction between two electrons on like (distinct) 
orbitals at the same 
lattice site is given by U(V).    
$H_0 = \sum_{\alpha} \sum_{\vec{k} \sigma} \left(\epsilon^0_{\alpha \vec{k} } - \mu \right) c^{\dagger}_{\vec k \alpha \sigma} c_{\vec k \alpha \sigma} $ is the dominant intra-orbital kinetic energy and gives rise to three decoupled energy bands at the Fermi level as shown in Fig. \ref{fs}.  Here, we make use of the following tight-binding parametrization of these energies: 
\begin{eqnarray}
&&\epsilon^0_{x(y)}(\vec k) = -2t \cos{k_{x(y)}} - 2t^{\perp}\cos{k_{y(x)}} \nonumber \\
&&\epsilon^0_{z}(\vec k) = -2t' \left( \cos{k_x} + \cos{k_y} \right) - 4 t'' \cos{k_x} \cos{k_y} 
\end{eqnarray}
where we take$(t,t^{\perp},t', t'', \mu) = (1.0,0.1,0.8,0.3, 1.0)$\cite{Kontani2008,Liu2008}.  The quantity $\delta H$ represents 
smaller terms such as longer range hopping and spin orbit coupling (SOC) which mix the distinct 
orbitals.   
It plays a relatively minor role in determining the superconducting transition temperature, 
However, $\delta H$ plays a crucial role in selecting a superconducting state which breaks time-reversal symmetry, as will be discussed below.  When $\delta H = 0$, the non-interacting  susceptibilities of the normal state are separate functions for each orbital:
\begin{equation}
\chi_{\alpha}(\vec q) = -\int \frac{d^2 k}{\left( 2 \pi \right)^2} \frac{f(\epsilon_{\alpha, \vec k + \vec q}) - f(\epsilon_{\alpha, \vec k})}{\epsilon_{\alpha, \vec k + \vec q} - \epsilon_{\alpha, \vec k}}
\end{equation}
where $f(\epsilon)$ is the Fermi  function.  
Since the quasi-two dimensional band is 
almost circular with a radius $k_f^{2d}$, its susceptibility is nearly  constant: $\chi_{z} \approx 1/4 \pi t'$ for $q < 2k_f^{2d}$. 
In contrast, the quasi-1D bands have susceptibilities that are peaked at   
 $\vec q_{x} = (2k_f^{1d}, \pi)$ 
  and $\vec q_{y} = (\pi, 2k_f^{1d})$ for the x and y orbitals, respectively.  It is  
the structure of $\chi_{x}$  and $\chi_{y}$ which gives rise to the incommensurate spin fluctuations in the material \cite{Sidis1999}.

Since the superconductivity in { \SRO} evolves out of a Fermi liquid and $T_c \ll E_f$, 
 it is reasonable to carry out a weak coupling analysis 
 which treats  the limit $U,V \ll  W$ where $W$ is the bandwidth.  
In this limit, superconductivity is the only instability of the Fermi liquid, and it can be treated in an asymptotically exact manner via a two-stage renormalization group analysis\cite{Raghu2010}.  In the first stage, high energy modes are perturbatively integrated out above an unphysical cutoff, and an effective particle-particle interaction in the Cooper channel is derived:
\begin{eqnarray}
\label{gamma}
\Gamma_s(\hat k, \hat q, \alpha) &=&U + U^2 \chi_{\alpha}(\hat k + \hat q) - 2V^2 \sum_{\beta \ne \alpha} \chi_{\beta} (\hat k - \hat q) \nonumber \\
\Gamma_t(\hat k, \hat q, \alpha) &=& -U^2 \chi_{\alpha}(\hat k - \hat q) - 2V^2 \sum_{\beta \ne \alpha} \chi_{\beta} (\hat k - \hat q)
\label{Gamma}
\end{eqnarray}
where 
$\Gamma_{a}(\hat k , \hat q, \alpha)$ ($a= s$ or $t$) is the effective interaction in the singlet(triplet) channel.  
In the second stage, the renormalization group flows of these effective interactions are computed and  
 the superconducting transition temperature is related to the energy scale at which an effective interaction grows to be of order 1.    
 Following
 this 
 prescription, one obtains
\begin{equation}
T_c \sim W e^{-1/\vert \lambda^{(a,\alpha)}_0 \vert}
\end{equation}
where $\lambda_0^{(a,\alpha)}$ is the most negative eigenvalue of the matrix
\begin{eqnarray}
g^{(a,\alpha)}_{\hat{k}, \hat{q}} = \sqrt{\frac{\bar{v}_f}{v_f(\hat k)}} \Gamma_{a}(\hat{k}, \hat{q}, \alpha)\sqrt{\frac{\bar{v}_f}{v_f(\hat q)}} 
\label{g}
\end{eqnarray}
with $\hat k$ and $\hat q$ constrained to lie on the Fermi surface of band $\alpha$.
The pair wavefunction in the superconducting state is  
proportional to the associated eigenfunction\cite{Raghu2010}.   
\begin{figure}
\includegraphics[width=3.0in]{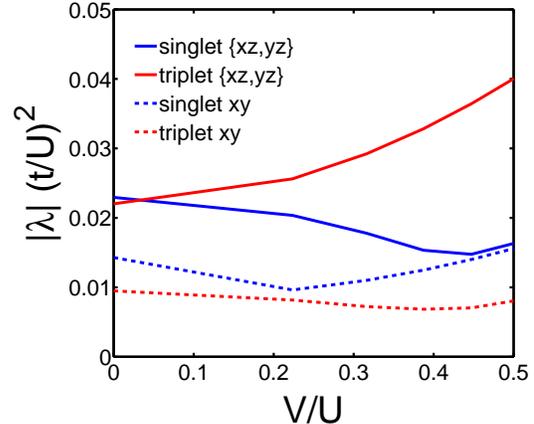}
\caption{(Color online) Pairing eigenvalues as a function of $V/U$ for the bandstructure parameters 
quoted in the text.  The strongest pairing strengths occur among the quasi-1D $\{\alpha, \beta\}$ bands.  There is a near 
degeneracy of the singlet and triplet eigenvalues for $V =0$, but with  $V > 0$, the quasi-1D
triplet state is the dominant superconducting configuration. }
\label{lambda}
\end{figure}

The values of $\lambda^{(a,\alpha)}$, obtained by numerical diagonalization,   
are  presented in Fig.  \ref{lambda}.  When $V=0$, the two dimensional 
z band has its dominant pairing instability in the singlet $d_{x^2-y^2}$ channel and a substantially lower pairing strength in the triplet p-wave channel.  By contrast, the pairing tendencies of the x and y bands are stronger, and exhibit a close competition between singlet and triplet pairing  
\footnote{A similar competition between singlet and triplet 
pairng is thought to occur in the quasi-one dimensional organic Bechgaard salts 
\cite{Shinagawa2007}. }.
\begin{figure}
\includegraphics[width=3.2in]{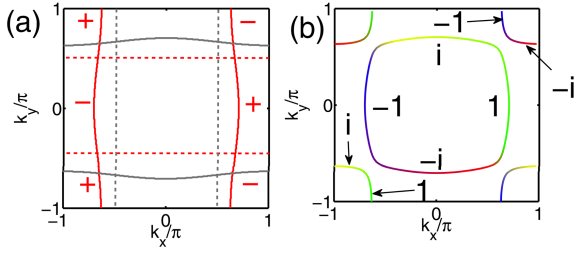}
\caption{(Color online) 
(a) In the absence of any band mixing, the triplet state within the xz orbital (red)  
 has $k_x$-wave 
symmetry and has line nodes near $k_y= \pi/2$ (dashed line).
The condensate on the yz orbital (grey) is related to the one 
shown here by a 90-degree rotation.  (b) The chiral state which 
results when small band mixing perturbations are taken into account. 
The relative phase factors on different portions of the Fermi surface are 
shown.  
   }
\label{wf}
\end{figure} 
When $V > 0$, only
triplet pairing in the quasi-one dimensional bands is enhanced.   
Since the solutions with weaker pairing strengths have {\em exponentially} smaller transition temperatures in the weak-coupling limit, the effect of subdominant orders is negligible.  
Thus, 
in the asymptotically weak coupling limit, {\it the dominant superconducting instability occurs in the quasi-one dimensional bands in the spin-triplet channel}.  
For $V > 0$, triplet pairing in the $x$ and $y$ bands is enhanced by virtual charge fluctuations occurring in the $z$ band. 
This conclusion is robust against a rather large range of quantitative changes of the band parameters\cite{Raghuinprep}.

The triplet pair wave-function in each band
\begin{equation}
\Psi_{\alpha}(\vec k)  = i \left[ \vec d_{\alpha}(\vec k) \cdot \vec \sigma  \sigma^y \right], \alpha = x,\ y
\end{equation}
  is specified by the  complex vector $ \vec d_\alpha(\vec k)$ in spin-space.
In general, the real and imaginary parts of $\vec d_\alpha$ are independent real vectors, and the net spin magnetization of band $\alpha$ is $\vec M_\alpha \propto \vec d_{\alpha}^*\times \vec d_{\alpha}$.  However, in weak coupling, 
only ``unitary states,'' {\it i.e.} states with $\vec M_\alpha = \vec 0$, need be considered, so $\vec d_\alpha$ can be expressed, up to a phase, as a real ``gap function'' times a real unit vector 
$\hat \Omega_\alpha$:
\be
\vec d_\alpha(\vec k) = \Delta_\alpha(\vec k)e^{i\theta_\alpha} \hat \Omega_\alpha.  
\ee
Figure \ref{wf}a shows the sign of $\Delta_x(\vec k) $  on the xz Fermi surface.  In addition to having odd parity, the wave function has two point nodes on the Fermi surface near $k_y = \pm \pi/2$ and is well approximated by   
$ \Delta_{x}(\vec k) \approx \Delta_0\sin{k_x} \cos{k_y}$.
 By symmetry, the 
 gap function  
 $\Delta_{y}(\vec k) \approx \Delta_0 \sin{k_y} \cos{k_x}$ on the yz Fermi surface.  
For $\delta H$=0, $\hat \Omega_x$, $\hat \Omega_y$, $\theta_x$ and $\theta_y$ are undetermined.

{\it Effect of small interactions -}
We now consider the effect of SOC and the hybridization among the different orbitals:
 \begin{eqnarray}
 \delta H &=&  \sum_{\vec k \sigma}
  \left[g(\vec k) c^{\dagger}_{\vec k , x, \sigma} c_{\vec k , y, \sigma} + {\rm H. C.}\right] \nonumber \\
&& + \eta \sum_{\alpha, \beta} \sum_{\sigma \sigma'} \sum_{\vec k} c^{\dagger}_{ \vec k, \alpha,  \sigma} c_{\vec k , \beta, \sigma'}  \vec \ell_{\alpha \beta} \cdot \vec \sigma_{\sigma \sigma'}
\end{eqnarray}
where $g(\vec k) = -2t'' \sin{k_x}\sin{k_y}$,  
 the second quantity above is the SOC, and the angular momentum operators are expressed in terms of the totally anti-symmetric tensor as $\ell^a_{\alpha \beta} = i \epsilon_{a \alpha \beta}$.  Recent electronic structure calculations 
have produced the estimates
$t''\approx 0.1t$ and $\eta \approx 0.1t $\cite{Haverkort2008, Liu2008}.  
  
There are several important qualitative effects of including these additional terms in the Hamiltonian: 
1)  Non-zero 
  $t^{\prime\prime}$ 
or $\eta$  pins  the relative phase, $\theta_x-\theta_y$, and relative orientation, $\hat\Omega_x\cdot\hat\Omega_y$ of the $d$ vectors. 2)  Non-zero $\eta$ defines a preferred ordering direction for $\Omega_a$.  
3)
  The nodes on the $xz$ and $yz$ Fermi surfaces are gapped, although the gap is parametrically small for small 
  $\delta H$.
To understand the role of 
$\delta H$ in selecting among the large number of possible ordered phases, it is simplest to consider the Landau free energy $\mathcal{F} $ to low order in powers of the order parameter which is a valid approximation near $T_c$.  
Since any order induced on the z band is slaved to the primary order on the x and y bands, we keep explicitly only $\alpha=x$ and $y$, in which case
 \begin{eqnarray}
&& \mathcal{F} = \sum_{\alpha}\left[\ r |\vec d_\alpha|^2 + u|\vec d_\alpha|^4+\gamma|\vec d^*_\alpha\times\vec d_\alpha|^2\ 
\right] \\
&& +
a_1\left[\ | d^z_x |^2+| d^z_y |^2 \right]
 + a_2\big[\ |d_x^x|^2+|d_y^y|^2\ \big] \nonumber\\
&&
  +v_1|\vec d_x|^2|\vec d_y|^2 + v_2 \vert \vec d_x \cdot \vec d_y \vert^2 
\nonumber\\
&&+ J_1 \left| \vec d^*_x \cdot \vec  d_y + \vec d^*_y  \cdot \vec d_x \right |^2+J_2 \left | \vec d^*_x \times \vec  d_y + \vec d^*_y  \times \vec d_x \right |^2
\nonumber
 \end{eqnarray}
where the terms in the first line survive the $\delta H\to 0$ limit, $a_j \sim {\cal O}( \eta^2/t^2)$, and $v_j$ and $J_j$ have contributions of order $(\eta/t)^2$ and $(t^{\prime\prime}/t)^2$.  To quadratic order in the order parameters and to order $(\eta/t)^2$, $(t^{\prime \prime}/t)^2$, this expression is the most general one consistent with symmetry, but for the quartic terms, in the interest of simplicity, we have assumed that the SOC is weaker than the band-mixing, and so have enforced spin rotational symmetry.  Since $|v_j| \ll u$, the terms on the third line are not qualitatively important.  
When the action is derived from any form of BCS  theory, it is possible to show that $\gamma$ and $ J_j>0$.  
  Thus, there are two possible phases depending on the sign and magnitude of $a_1$:
  (a) For $a_1 >$ Min$[0, a_2]$, there are  time-reversal symmetry preserving 
    ``B'' phase states (analogous to the B-phase in $^3$He) in which
  $\hat \Omega_x \cdot \hat \Omega_y=\hat \Omega_x\cdot\hat z = \hat \Omega_y\cdot\hat z = 0$ and 
    $ \theta_x=\theta_y$.  Depending on the sign of $a_2$,  either $\hat\Omega_x = \hat x$ or $\hat\Omega_x = \hat y$.   (b) For $a_1 >$ Min$[0, a_2]$,  there is a ``chiral p+ip'' state (analogous to the A phase in He3) with $\hat \Omega_x =  \hat \Omega_y=\hat z$
    and 
    $\theta_x -\theta_y \pm \pi/2$.  All other configurations have a higher Free energy and can be neglected.  
  
  The parameters $a_j$ can be related to differences of susceptibilities of the non-interacting system, and so can be computed directly from the assumed band-structure.  For the stated parameters, we find that $a_1 <$ Min$[0,a_2]$, so the chiral state is preferred.  However, the balance is delicate, and 
     this conclusion is not robust against small changes to the model.

Next, we address the  
fate of the gap nodes  
 when $\delta H \ne 0$.  
 We have studied   the  BdG Hamiltonian for 
  the
  chiral state using the gap functions derived in the previous section.  
 Generically, the resulting state is nodeless.  However, although the nodes are not topologically stable, 
they are parametrically small:   where a gap node occurs for $\delta H =0$, the induced gap is
$\sim \Delta_0 \left[ {\cal O}(t''^2/t^2)+{\cal O}(\eta^2/t^2)\right] $. 
The energy scale of these gap minima is therefore two orders of magnitude smaller 
than the transition temperature.

{\it Topological properties and edge currents -}
Next, we consider the topological properties of the system 
assuming that $a_1 <$ Min$[0,a_2]$, so the chiral state is preferred.  
The BdG Hamiltonian for the quasiparticle excitations in the superconducting state can then be expressed in terms of Anderson pseudo-spins as 
\begin{equation}
H_{BdG} =
\sum_{\nu,\vec k}  \Psi^{\dagger}_{\nu \vec k }\left[ \vec \delta_{\nu}(\vec k) \cdot \vec \tau 
\right] \Psi_{\nu \vec  k }.
\end{equation}
where $\Psi_{\nu \vec k}$ are Nambu spinors, $\nu=\alpha$, $\beta$ runs over the two quasi-1D bands,  $\vec \tau$ are the Pauli matrices, and
the pseudo-Zeeman field is
\begin{equation}
\vec\delta_{\nu}(\vec k) = \left({\rm Re}[ \Delta_{\nu}(\vec k)],{\rm Im}[ \Delta_{\nu}(\vec k)], \epsilon_{\nu}( \vec k)- \mu \right).
\end{equation}
For the chiral p-wave state, the pseudospin has the form of a skyrmion in momentum space: it points along the $-\hat z(+\hat z)$-direction inside(outside) the Fermi surface, and on the Fermi surface, it lies in plane, winding by $2 \pi$ around the Fermi surface.  The topological properties of the chiral state come from the integer skyrmion number 
\begin{equation} 
\mathcal N_{\nu} = \frac{1}{4 \pi |\vec \delta_\nu(\vec k)|^3} \int d^2 k  \vec  \delta_{\nu} \cdot \left( \partial_x \vec \delta_{\nu} \times \partial_y \vec \delta_{\nu}\right)
\end{equation}
where   
$|\vec \delta_\nu(\vec k)|=\sqrt{[\epsilon_{\nu}( \vec k) - \mu]^2+|\Delta_\nu(\vec k)|^2}\ $. 
The net number of chiral quasiparticle modes at the edge of the superconductor is given by the skyrmion number, and so long as $\mathcal N_{\nu} \ne 0$,  
these modes are topologically protected, and  
cannot be localized by backscattering.

As $\mathcal N_{\nu}$ is an integer, small changes in the spectrum do not affect it.  However, it is odd under $\Delta_\nu \to \Delta_\nu^*$ ({\it i.e.}  upon transforming  $p_x + i p_y \rightarrow p_x - i p_y$) or under a particle-hole transformation, $\epsilon_{\nu} - \mu \rightarrow \mu - \epsilon_{\nu}$.  As can be seen in Fig. 1, hybridization between the two quasi-1D bands  results in the closed $\alpha$ and $\beta$ Fermi surfaces, the former electron-like and the latter hole-like.  Consequently, in a chiral $p_x+ip_y$ state, $\mathcal N_{\alpha} =-\mathcal N_{\beta}$, or in otherwords the net skyrmion number is zero so
 it is not a topological superconductor!  
The chiral edge modes along the boundary of the superconductor and along domain 
walls are 
not protected: in the presence of disorder or interactions that scatter a pair from one Fermi surface to the other, the  counter-propagating edge modes from the two bands are localized 
\cite{Raghuinprep}.

The existence of  edge currents in a chiral p-wave superconductor follows from symmetry, but whether or not it is large enough to detect is still an issue.  There is a bulk contribution to the edge current which originates from the multi-component nature 
of the order parameter,
and one from the 
chiral quasiparticle modes near the edge of the system.  We will address the question of how the highly non-circular band structure of the quasi 1D bands affects the bulk contribution in a future publication \cite{Raghuinprep}.  Since the
chiral quasiparticle modes travel with 
velocity $v = v_f ( \Delta/E_f)$, they would, by themselves, make a readily detectable 
contribution $\mathcal O(\Delta/E_f)$ to the edge 
current, were they not localized. 

 {\it Discussion - }
 It follows from general arguments\cite{Agterberg1997} that  near T$_c$,
superconductivity can arise either in the $ \{ \alpha, \beta \}$ pockets or in the   $\gamma$ pocket; 
below T$_c$, superconductivity is induced in the subdominant Fermi surfaces via a proximity effect 
\begin{equation}
\delta H_{prox} =J' \sum_{\nu \ne \nu'} c^{\dagger}_{i \nu \uparrow }c^{\dagger}_{i \nu \downarrow} c_{i \nu' \downarrow} c_{i \nu' \uparrow},
\end{equation}
with $J' \ll U$\cite{Agterberg1997}.  Due to the weakness of this proximity effect, it can be expected that for a range of temperatures $J'(\Delta/E_F) \ll T < T_c$, superconductivity is present essentially only on the dominant portions of the Fermi surfaces.  Nonetheless, in the present case,  
this coupling   produces a small gap on the xy Fermi surface, which then
adds to the skyrmion number; while the energy scales involved are likely too small to affect the results of any practical experiment, ultimately the proximity effect restores the topological character of the chiral superconducting state.

As we have shown, it is unavoidable, given the band-structure of \SRO, and assuming the interactions are weak, that the superconductivity arises primarily on the quasi-1D bands.  
Within this framework, 
there is a natural 
explanation for the surprising absence of superconductivity in the closely 
related bilayer compound Sr$_3$Ru$_2$O$_7$; 
the bilayer splitting in Sr$_3$Ru$_2$O$_7$ primarily affects the quasi-1D 
bands, leaving the 2D $d_{xy}$ band essentially unchanged.  If pairing occurred primarily on the quasi-2d 
band, 
both compounds ought to 
be similarly superconducting.  (An 
 analysis of superconductivity in the bilayer system will be presented in a separate publication\cite{Raghuinprep}). 

Admittedly, we cannot rule out the possibility that strong coupling effects will change this conclusion.  However, the absence of experimentally detectable edge currents  is difficult to reconcile with a state which primarily involves the 2D bands.  
Further analysis of the experimental evidence concerning this issue will be postponed to a future paper\cite{Raghuinprep}.  
We will also discuss the nature of the  collective modes which reflect the near independence of the superconducing state on each of the quasi-1d bands,
especially near the transition temperature.

We are grateful to D. Agterberg, S.-B. Chung, C. Kallin, A.P. MacKenzie, K. Moler, D. Scalapino, and S.-C. Zhang for helpful 
discussions.  This work is supported in part by NSF grant number DMR-0758356 (SR and SAK) 
and by the Department of Energy Grant DE-AC02-76SF00515 (AK) at 
Stanford University.

\bibliography{SrRuO214}
\end{document}